\documentclass[prl,aps,amssymb,twocolumn,superscriptaddress,notitlepage,nobalancelastpage]{revtex4}
\usepackage{amsmath}
\usepackage{amsthm}
\usepackage{amsfonts}
\usepackage{listings}
\usepackage{enumerate}
\usepackage{latexsym}
\usepackage{psfrag}
\usepackage{bm}
\usepackage{graphicx}
\usepackage[caption=false]{subfig}
\usepackage{array}
\usepackage{color}
\usepackage[normalem]{ulem}
\usepackage[colorlinks=true ,citecolor=blue]{hyperref}

\usepackage{esint}

\usepackage{graphicx}
\usepackage{float}
\usepackage{braket}

\begin{document}
\title{Tutorial: Computing topological invariants in two-dimensional photonic crystals}
\author{Mar\'ia Blanco de Paz}
\author{Chiara Devescovi}
\affiliation{Donostia International Physics Center, 20018 Donostia-San Sebastian, Spain}
\author{Geza Giedke}
\author{Juan Jos\'e Saenz}
\author{Maia G. Vergniory}
\affiliation{Donostia International Physics Center, 20018 Donostia-San Sebastian, Spain}
\affiliation{IKERBASQUE, Basque Foundation for Science, Maria Diaz de Haro 3, 48013 Bilbao, Spain}
\author{Barry Bradlyn}
\affiliation{Department of Physics and Institute for Condensed Matter Theory, University of Illinois at Urbana-Champaign, Urbana, IL, 61801-3080, USA}
\author{Dario Bercioux}
\author{Aitzol Garc\'{i}a-Etxarri}
\affiliation{Donostia International Physics Center, 20018 Donostia-San Sebastian, Spain}
\affiliation{IKERBASQUE, Basque Foundation for Science, Maria Diaz de Haro 3, 48013 Bilbao, Spain}
\date{\today}

%%%%%%%%%%%%%%%%%%%%%%%%%%%%%%%%%%%%%%%%%%%%%%%%%%%%%%%%%%%%%%%%%%%%%%%%%%%%%%%%%%%%%
\begin{abstract}
The field of topological photonics emerged as one of the most promising areas for applications in transformative technologies: possible applications are in topological lasers or quantum optics interfaces.
Nevertheless, efficient and simple methods for diagnosing the topology of optical systems remain elusive for an important part of the community. In this tutorial, we provide a summary of numerical methods to calculate topological invariants emerging from the propagation of light in photonic crystals.
We first describe the fundamental properties of wave propagation in lattices with a space-dependent periodic electric permittivity. Next, we provide an introduction to topological invariants; proposing an optimal strategy to calculate them through the numerical evaluation of Maxwell's equation in a discretized reciprocal space. 
Finally, we will complement the tutorial with a few practical examples of photonic crystal systems showing different topological properties, such as photonic valley-Chern insulators, photonic crystals presenting an ``obstructed atomic limit'', photonic systems supporting fragile topology and finally photonic Chern insulators, where we  also periodically modulated the magnetic permeability. 
\end{abstract}
%%%%%%%%%%%%%%%%%%%%%%%%%%%%%%%%%%%%%%%%%%%%%
%%%%%%%%%%%%%%%%%%%%%%%%%%%%%%%%%%%%%%%%%%%%%%%%%%%%%%%%%%%%%%%%%%%%%%%%%%%%%%%%%%%%%

\maketitle

%%%%%%%%%%%%%%%%%%%%%%%%%%%%%%%%%%%%%%%%
%%%%%%%%%%%%%%%%%%%%%%%%%%%%%%%%%%%%%%%%%%%%%%%%%%%%%%%%%%%%%%%%%%%%%%%%%%%%%%%%%%%%%

\section{Introduction}

Over the past decades, the interest in topologically nontrivial materials increased dramatically, giving rise to the development of topological insulators, semi-metals, and superconductors, among others. 
Thouless $et~al.$~\cite{tknn} in 1982 proposed the first topological characterization of the integer quantum Hall effect (QHE) in a material in terms of the Chern number. Following these ideas, several types of topologically nontrivial insulators have been reported~\cite{Haldane1988,Kane04,fukanemele,bernevig2006quantum,konig2007quantum,Xia09}. 
These materials, besides presenting anomalous bulk response functions~\cite{qi2008topological,essin2009magnetoelectric,teo2010topological}, are particularly relevant because they feature topologically protected states such as surface, edge, hinge and corner modes~\cite{hotis,khalaf,benalcazar2017quantized,benalcazar2017electric,benalcazar2018quantization,fangrotational,hingeprep,Kempkes2019,hotis2}.

Despite the fact that most of the crucial concepts in band topology were first developed in condensed matter physics, many of them were soon translated to the propagation of electromagnetic waves in photonic crystals~\cite{Joannopoulos1997}. In the recent past, bosonic analogs of the QHE ~\cite{wang2008reflection, wang2009observation}, quantum spin-Hall effect (QSHE)~\cite{khanikaev2013photonic,hafezi2011robust, umucalilar2011artificial}, quantum valley-Hall effect~\cite{ma2016all}, Floquet topological insulators~\cite{fang2012realizing,maczewsky2017observation, mukherjee2017experimental,rechtsman2013photonic}, mirror-Chern and quadrupole insulating~\cite{2019arXiv190107154Z,hassan2018corner, ota2018photonic, xie2018second} systems have been reported. Moreover, the emergence of topological effects has also been extended to the propagation of other waves such as airborne acoustics~\cite{he2016acoustic,Peri2019}, exciton--polaritons~\cite{klembt2018exciton}, phonons~\cite{Ssstrunk2016,Liu2017,Jiang2018}, and spin waves~\cite{RoldnMolina2016,Yamamoto2019}. 

The classification of band insulators is fully determined by different topological invariants~\cite{KitaevClassify,Shiozaki2017,Po2017,NaturePaper,schnyder2008classification}. In  systems without time-reversal symmetry (TRS), the Chern number $C$, characterizes the topological properties of an isolated band or a group of connected bands. When TRS is preserved, although the Chern number is strictly zero, other topological invariants become relevant, such as the $\mathbb{Z}_2$~index. Moreover, in certain systems, time-reversal relates pairs of high symmetry points in reciprocal space. For instance, in a triangular lattice, the TRS operator transforms $K$ point/valley into $K'$ point/valley and \emph{vice versa}~\cite{Suzuura2002}. In such cases, the valley can be approximately treated as a new degree of freedom, and other topological numbers, such as the valley Chern number $C_{v}$~\cite{fang2012realizing,rechtsman2013photonic,hafeziimaging} can become relevant in the description of the topological properties of a material. In these scenarios, although the total Chern number $C$ vanishes, certain topological effects determined by $C_{v}$ may arise in specific  valleys. 

The aim of this tutorial article is to provide an overview of the description of the computation of the most relevant topological invariants to the propagation of light through two-dimensional (2D) linear photonic crystals. The procedures described in this article can be applied to any 2D lattice structure; particularly, we provide strategies to compute the total Chern number, the valley-Chern number and the Wilson loop --- from the latter we can infer the value of the $\mathbb{Z}_2$ topological index. We combine traditional approaches to calculate Chern numbers, by integrating the Berry curvature over regions of the reciprocal space with another important approach in the study of topological phases consisting in the evaluation of Wilson loops ---  giving information about the evolution of the Wannier centers around closed loops in the Brillouin zone (BZ)~\cite{Lu2016,Neupert2018,Alexandradinata2019}. 

Topological photonics are interesting from at least two
perspectives. First, they are able to shape and guide the
electromagnetic field, allowing technologically important effects
like the robust unidirectional propagation of light~\cite{khanikaev2013photonic,Lodahl_2017,ozawa2018topological}. Second, the flexibility of photonic crystal structures and metamaterials makes possible the exploration of topological physics in bosonic systems, including different topological phases and the interplay of topology with non-Hermitian
effects~\cite{Martinez_Alvarez_2018,ozawa2018topological,Zhao2019}

The quantum aspects of topological photonics become especially
important when topological structures are coupled to quantum
emitters. The emitters then probe a highly structured environment
which can lead to a rich variety of new effects, the physics of which
has only begun to be studied~\cite{ozawa2018topological,Bello2019}.
On one hand, the quantum emitters can locally probe the photonic
structure, extracting information about the local density of states or
about the topological properties. They can also be used to excite specific,
topologically interesting modes with single or multiple photons, for
example using topologically protected edge states to transmit quantum
information or entanglement. On the other hand, the structured environment can mediate interactions between quantum emitters. These interactions can be long-range,
reflecting the non-local character of topological bands or even
induce non-trivial topological properties in the system of quantum
emitters, where they may give rise to interesting topological phases including quantum many-body effects, that are not easily obtained for (non-interacting) photons. 

For all these applications, a precise knowledge of the mode structure
and the topological properties of the photonic structured bath is
essential. This tutorial provides a toolbox that will prove
useful to study the interaction of concrete photonic structures
with quantum emitters such as atoms or defect centers.

This article is structured as follows: In the first section, we introduce the eigenvalue equation for the propagation of light in photonic crystals. Next, we review the analytic formalism in the continuum generally used to compute topological invariants. Following that, we focus on the discretization of the Brillouin zone for numerical calculations paying attention to the boundary constraints needed to obtain reliable results. Finally, we conclude this tutorial article presenting various examples of photonic systems with non-trivial topological properties: A valley Chern insulator, lattices presenting an photonic ``obstructed atomic limit'', a system with a group of bands that posses fragile topology, and a Chern insulator.  \\

%%%%%%%%%%%%%%%%%%%%%%%%%%%%%%%%%%%%%%%%%%%%%%%%%%%%%%%%%%%%%%%%%%%%%%%%%%%%%%%%%%%%%
%%%%%%%%%%%%%%%%%%%%%%%%%%%%%%%%%%%%%%%%%%%%%%%%%%%%%%%%%%%%%%%%%%%%%%%%%%%%%%%%%%%%%

\section{Macroscopic Maxwell's equations in periodic media}\label{Maxwell} 

Electromagnetic wave propagation in photonic crystals is governed by the macroscopic Maxwell's equations. Assuming that the materials under consideration are linear, isotropic and dispersionless, the source-free Maxwell's equations can be expressed in the frequency domain as 
%
%
%%%%%%%%%%%%
\begin{subequations}\label{Maxwell:eqs}
\begin{align} 
   & \nabla \cdot \mathbf{H}(\mathbf{r})  = 0 \label{Maxwell1} \\
    & \nabla \times \mathbf{E}(\mathbf{r}) -i\omega  \mathbf{H}(\mathbf{r}) = 0\label{Maxwell2}\\
    & \nabla \cdot \left[\epsilon(\mathbf{r})\mathbf{E}(\mathbf{r})\right] = 0 \label{Maxwell3} \\
& \nabla \times \mathbf{H}(\mathbf{r}) + i\omega\epsilon_0\epsilon(\mathbf{r})\mathbf{E}(\mathbf{r})= 0  \label{Maxwell4}
\end{align}
\end{subequations}
%%%%%%%%%%%%
%
%
where we are assuming harmonic $e^{-i\omega t}$ time dependence for the electric and magnetic fields. In the previous equations, $\mathbf{E}(\mathbf{r})$ and $\mathbf{H}(\mathbf{r})$ are the electric and magnetic field vectors, respectively. In this first section of the tutorial, we will work with non-magnetic materials. Thus, $\mu=1$ and $B=H/\mu_0$. Following Ref.~\cite{photoniccrystals}, through the appropriate algebraic manipulation, Eqs.~(\ref{Maxwell1}-\ref{Maxwell4}) are decoupled and expressed as two separate eigenvalue problems in terms of either the magnetic field $\mathbf{H}(\mathbf{r})$ or the electric field $\mathbf{E}(\mathbf{r})$:
%
%
%%%%%%%%%%%%
\begin{subequations}\label{eq:master:Gen}
\begin{align}\label{eq:master}
\nabla \times \left[ {\frac {1}{\epsilon(\mathbf{r})}} \nabla \times \mathbf{H}(\mathbf{r}) \right] & = \left( {\frac {\omega}{c}} \right)^{2} ~ \mathbf{H}(\mathbf{r}), \\
\nabla \times  \nabla \times \mathbf{E}(\mathbf{r}) & = \left( {\frac {\omega}{c}} \right)^{2} ~ \epsilon(\mathbf{r})\mathbf{E}(\mathbf{r})\,,\label{eq:masterE}
\end{align}
\end{subequations}
%%%%%%%%%%%%
%
%
where we have used the definition of speed of light  $c=1/\sqrt{\epsilon_0 \mu_0}$. 
To determine the eigenmodes of a particular system the first equation is commonly used since can be solved as a Hermitian eigenvalue problem unlike the second which is another type of equation, the so called \emph{generalized} eigenvalue equation, because of the presence of the function $\epsilon(\mathbf{r})$ on the right-hand side. Therefore, we usually address the first of these equations to compute the eigensolutions of the magnetic field and then recover the electric field from the magnetic field~\cite{sakoda,photoniccrystals}.
Many commercial and open source software packages exist to solve this eigenvalue problem, such as COMSOL Multiphysics \cite{comsol} and MIT Photonic Bands (MPB) \cite{mpb}, the most commonly used in the field of photonics.
All the calculations presented in this tutorial have been developed using the latter. \\

For light-matter interactions in three-dimensional (3D) structures, these equations must be solved as a fully vectorial problem. Nevertheless, in a 2D systems, mirror reflection symmetry always allows  to classify the solutions to Eq.~\eqref{eq:master:Gen} into even and odd solutions with respect to it. In general, for any 2D photonic crystals, assuming that $\hat{z}$ is the invariant direction of the system, the eigensolutions can be either polarized as $(E_x,E_y,H_z)$ (mirror even) or $(H_x,H_y,E_z)$ (mirror odd). These two polarizations are commonly known as transverse electric (TE) or transverse magnetic (TM) modes, respectively.

For TM modes, the magnetic field is confined to the $xy$-plane, therefore the  non-zero components of the magnetic field are ${H}_x(\mathbf{r}),{H}_y(\mathbf{r})$. Most importantly, the electric field becomes a scalar function, being ${E}_z(\mathbf{r})$ the only nonzero vector component. Similarly, for TE modes, ${H}_z(\mathbf{r})$ is the only nonzero component of the magnetic field.
Consequently, Eqs.~\eqref{eq:master} and \eqref{eq:masterE} can be rewritten for 2D photonic crystals as two scalar equations, one for each of polarization:
%
%
%%%%%%%%%%%%%%%%
\begin{subequations}
\begin{align}
\nabla \times \nabla \times {E}_z(\mathbf{r}) & = \left( {\frac {\omega}{c}} \right)^{2} ~ \epsilon(\mathbf{r}){E}_z(\mathbf{r}) \label{eq:TM}\\  &\mathrm{for~the~TM~modes}\nonumber \\
\nabla \times \left[ {\frac {1}{\epsilon(\mathbf{r})}} \nabla\times {H}_z(\mathbf{r}) \right] & = \left( {\frac {\omega}{c}} \right)^{2} ~ {H}_z(\mathbf{r})\label{eq:TE}\\
&\mathrm{for~the~TE~modes} \nonumber 
\end{align}
\end{subequations}
%%%%%%%%%%%%%%%%%
%
%

We can now recast the differential equation for the TE modes into an eigenvalue problem by defining the following differential operator:
%
%
%%%%%%%%%%%%%%%%%
\begin{equation}
    \hat{\Theta}[\#]=  \nabla \times \left({\frac {1}{\epsilon(\mathbf{r})}} \nabla\times\# \right) 
\end{equation} 
%%%%%%%%%%%%%%%%%
%
%
which is a linear and Hermitian~\cite{sakoda,photoniccrystals}. 
Equation~\eqref{eq:TE} for the TE modes now is reduced to
%
%
%%%%%%%%%%%
\begin{equation}\label{eq:TETM}
        \hat{\Theta}~ {H}_z(\mathbf{r})  = \left( {\frac {\omega}{c}} \right)^{2} ~ {H}_z(\mathbf{r})\,,
\end{equation}
%%%%%%%%%%%
%
%
where $(\omega/c)^2$ is the eigenvalue and $H_z(\mathbf{r})$ is the eigenfunction. 
A solution for the TE modes can be easily obtained exploiting the periodicity of the electric permittivity $\epsilon(\mathbf{r})$ ---  since we can make use of the Bloch's theorem.  Thus, in general, the eigensolutions of Eq.~\eqref{eq:TE} can be expressed as Bloch states labeled with a wave vector $\mathbf{k}$:
%
%%%%%%%%%%
\begin{equation}\label{eq:1}
    H_{z,\mathbf{k}}(\mathbf{r})= e^{i\mathbf{k}\cdot\mathbf{r}} u_{\mathbf{k}}(\mathbf{r})
\end{equation}
%%%%%%%%%%
%
%
where $u_{\mathbf{k}}(\mathbf{r})$ are periodic functions of the lattice, \emph{i.e.} $u_{\mathbf{k}}(\mathbf{r})=u_{\mathbf{k}}(\mathbf{r}+\mathbf{R})$, and $\mathbf{R}$ is the lattice periodicity.\\
Please note that  Eq.~\eqref{eq:TETM} is formally equivalent to the stationary Schr\"odinger equation
%
%
%%%%%%%%%%
\begin{equation}\label{eq:Schro}
        \hat{H} ~\Psi(\mathbf{r})  = E ~ \Psi(\mathbf{r})\,. 
\end{equation}
%%%%%%%%%
%
%
This formal equivalence and the use of the Bloch theorem in Eq.~\eqref{eq:1}, allow us to apply many of the concepts and theorems used in condensed matter physics to the study of light propagation through photonic crystals, being aware about the bosonic character of photons. Nevertheless, it is well known that the main concepts in electronic band topology, within the single-particle approximation, can also emerge similarly in the propagation of light in specific periodic photonic structures. These phenomena, require the proper characterization of either electronic or photonic crystals through the evaluation of the  topological invariants. Although there are a vast literature and computational resources for the computation of such quantities in condensed-matter systems such as Z2PACK~\cite{z2pack}, WANNIER90~\cite{Marzari2012} among others, literature and tools about the computation of these topological invariants for classical waves are still very sparse.

In the following sections, we proceed to describe the details of such calculations, focusing particularly on the computation of topological invariants in 2D photonic crystals. Our results have been developed using the MPB software package~\cite{mpb}. Nonetheless, the provided numerical methods apply easily also using other tools for solving the differential problem in Eq.~\eqref{eq:master:Gen}. 

%%%%%%%%%%%%%%%%%%%%%%%%%%%%%%%%%%%%%%%%%%%%%%%%%%%%%%%%%%%%%%%%%%%%%%%%%%%%%%%%%%%%%
%%%%%%%%%%%%%%%%%%%%%%%%%%%%%%%%%%%%%%%%%%%%%%%%%%%%%%%%%%%%%%%%%%%%%%%%%%%%%%%%%%%%%

\section{Fundamentals of topology, the continuum limit}\label{TopologicalInvariantsContinous}

In this section we present a concise review of the main topological invariants that are used throughout this tutorial article. Many specialized articles and books work out these concepts in a more extended and rigorous approach. For readers interested to explore this topic in more depth, we suggest these two references~\cite{vanderbilt2018,ozawa2018topological}.

In general, given a quantum or a classical periodic differential operator, its spectrum is arranged in a sequence of bands and gaps. 
Many of the physical properties of electronic materials and photonic crystals can be inferred from the intensity distribution of the Bloch states and their bulk spectrum. Nevertheless, these quantities cannot tell us anything about the evolution of eigenmodes' phase in $k$-space. This is a piece of essential information to understand and  predict certain emergent topological phenomena such as the QHE or QSHE, among others. 
In an electronic system, the topological properties are referred to the bands below the Fermi level. For classical wave systems, the concept of Fermi level does not exist. However, it is possible to address individually the bands and gaps of the system by tuning the frequency of the photons exciting the lattice. Therefore, we are interested in characterizing the topological properties associated to gaps, analyzing isolated bands or group of bands above and below the gap.

Given an group of $M$ bands,  we are interested in the situation in which there is an  energy band with index $n<M$, that is $\omega_{n-1}(\mathbf{k})<\omega_n(\mathbf{k})<\omega_{n+1}(\mathbf{k}), ~~\forall~ \mathbf{k}\in \mathrm{BZ}$. Further, we assume that this band is not degenerate. 
The  $n$-th band can be characterized by the study of the so-called Berry-Pancharatnam-Zak phase~\cite{Berry1984,Pancharatnam1956,vanderbilt2018,Zakphase}. This is a geometric phase that the eigenstate of the $n$-th band acquires during an adiabatic evolution along a closed path $\gamma$ in reciprocal space. The Berry-Pancharatnam-Zak phase is obtained by the following closed path integral:
%
%
%%%%%%%%
\begin{equation}\label{3.1}
   \phi_n= \oint \mathbf{A}_n(\mathbf{k})\cdot d\mathbf{k}\,,
\end{equation}
%%%%%%%%
%
%
\noindent where $A_n(\mathbf{k})$ --- also known as Berry connection --- is defined as
%
%
%%%%%%%%
\begin{equation}\label{3.2}
    \mathbf{A}_n(\mathbf{k})\equiv i \braket{u_{n,\mathbf{k}}|\nabla_{\mathbf{k}}|u_{n,\mathbf{k}}}\,.
\end{equation}
%%%%%%%%
%
%
The Berry-Pancharatnam-Zak phase is gauge invariant up to multiples of $2\pi$ and depends only on the geometry of the path  $\gamma$, whereas the Berry connection is  gauge-dependent. An important gauge independent quantity is the Berry curvature $\bm{\Omega}_n (\mathbf{k})$: 
%
%
%%%%%%%%
\begin{equation}\label{3.3}
   \bm{\Omega}_n (\mathbf{k})= \bm{\nabla}_{\mathbf{k}} \times \mathbf{A}_n(\mathbf{k})\,.
\end{equation}
%%%%%%%%
%
%
We can express $\phi_n$ as a function of $\bm{\Omega}_n(\mathbf{k})$ applying Stoke's theorem to Eq.~\eqref{3.1}, 
%
%
%%%%%%%%%%%%
\begin{equation}\label{3.4}
\phi_n=\oint \mathbf{A}_n(\mathbf{k}) \cdot d \mathbf{k}=\iint_{\Gamma} [\bm{\nabla}_{\mathbf{k}} \times\mathbf{A}_n(\mathbf{k})]\,d\mathbf{s}\equiv\iint_{\Gamma}\bm{\Omega}_n(\mathbf{k}) d\mathbf{s}
\end{equation}
%%%%%%%%%%%%
%
%
where $\Gamma$ is a surface bounded by the closed path $\gamma$ along which the integral in Eq.~\eqref{3.1} is evaluated.

If the surface $\Gamma$ is a closed manifold, the boundary term vanishes, but the indeterminacy of the boundary term modulo $2\pi$ manifests itself in the Chern theorem~\cite{Chern1946}; this states that the integral of the Berry curvature over a closed manifold is quantized in units of $2\pi$. This number is the so-called Chern number, and is essential for understanding various quantization effects when TRS is broken, such as as the QHE~\cite{firstchern}.
The Chern number for a two-dimensional manifold is defined as:
%
%
%%%%%%%%%
\begin{equation}\label{3.5}
    C_n= \frac{1}{2\pi} \oiint_\mathrm{BZ}  \bm{\Omega}_n (\mathbf{k})  \,d^2 \textrm{k}~~~\textrm{with}~~C_n\in\mathbb{Z},
\end{equation}
%%%%%%%%%
%
% 
where the closed surface $\Gamma$ coincides with the torus defining the first BZ. 

When TRS is preserved, the Chern number is zero. However, in the case of systems with a valley (or other ``internal'') degree of freedom in the BZ, we can partition the integral in Eq.~\eqref{3.5} to isolate the contribution from each single valley and introduce the so-called valley-Chern number~\cite{Ezawa2013}. Each valley-Chern number can be different from zero even if TRS is conserved, but the sum of all the valley-Chern numbers, being equal to $C_n$, has to be zero.

Aside from valley-Chern systems, other materials preserving TRS can also have topological properties. This is the case of systems supporting QSHE~\cite{Kane04} and fragile phases ~\cite{Fragile2017,comment,Slager2018}. In those cases, studying the eigenvalues of the Wilson loop operator helps to determine the topological nature of the system. This operator is obtained considering a non-Abelian generalization of the Berry-Pancharatnam-Zak phase. Here, we assume that the $n$-th band can be degenerate with one or more bands. For $N$ connected bands, the Wilson loop operator is derived from the non-Abelian Berry connection:
%
%
%%%%%%%%%%%%
\begin{equation}\label{3.6}
\mathbf{A}_{m,n}(\mathbf{k})=i \langle u_{\mathbf{k},m}|\nabla_{\mathbf{k}}|u_{\mathbf{k},n}\rangle, ~~~n,m=1,\ldots N\,;
\end{equation}
%%%%%%%%%%%%
%
%
it is the unitary matrix
%
%
%%%%%%%%%%%%
\begin{equation}\label{3.7}
W(\ell)= \mathcal{P}\exp\left\{-i\int_\ell d\mathbf{l}\cdot \mathbf{A}(\mathbf{k})\right\}\,,
\end{equation}
%%%%%%%%%%%%
%
%
where $\ell$ is a loop in  momentum space and $\mathcal{P}$ denotes a path ordering of the exponential. 
It can be shown that for a path $\ell$ described by a straight line through the Brillouin zone, the eigenvalues of the Wilson loop correspond to the expectation value of the position operator in the remaining real-space direction modulo $2\pi$~\cite{vanderbilt2018,Neupert2018}. The spectrum of the Wilson loop is thus adiabatically deformable to the set of centers of localized Wannier functions describing the group of bands. For trivial bands, the Wannier states are maximally localized, while for systems that posses non-trivial topology, the Wannier states are delocalized. Thus, the Wilson loop gives us information about the topological character of a band or a group of bands.
In two dimensions, one of the momenta $k_1$ defines the integration variable of the closed path $\ell$ in Eq.~\eqref{3.7}, while the other momentum $k_2$ is a free parameter characterizing the Wilson loop.  For a topologically trivial system, the eigenvalues along the momentum $k_2$ are adiabatically deformable to a constant value, reflecting that the Wannier functions are localized. For a non-trivial topological system, the eigenvalues may wind as a function of the momentum $k_2$, meaning that the eigenvalues of the Wilson loop present a variation of $2\pi n,\,\, n\in\mathbb{Z}$ along the BZ (cf Sec.~\ref{Chernexample})--- this corresponds to the delocalization of the Wannier functions.
Additionally, from the slope and the number $n$ of windings we can extract important information: The sign of the slope will tell us about the sign of the topological number, whereas the number of windings will tell us its absolute value.
Another possible phase is the \emph{photonic} obstructed atomic limit (OAL), which presents non-winding values of the Wilson loop. This phase supports a maximally localized Wannier representation, but differs from the trivial phase in that the Wannier centers are not located at the origin of the real space unit cell. In analogy with an electronic system, in a photonic OAL 
the Wannier centers are not located at the position where the photonic ``atoms''-- the collection of dielectric objects in the unit cell--sit\cite{Joannopoulos1997}.

Moreover, there exists a recently discovered class of topological indices referred to as ``fragile.'' The simplest of these display Wilson loop eigenvalues which wind with opposite slopes, indicating that although the total Chern number is equal to zero the system presents topological features similar to those of $\mathbb{Z}_2$ insulators. The special feature of this invariant is that if a trivial band is added in the non-Abelian Berry connection~\eqref{3.6}, then the eigenvalues of this enlarged system behave as in the case of the OAL~\cite{Alexandradinata2019}, losing the topological character. 

Additionally, the Wilson loops are very useful to identify $\mathbb{Z}_2$-insulators. Note that $\mathbb{Z}_2$ topology in photonic systems requires of duality symmetry and crystal symmetries as a proxy for fermionic TRS~\cite{khanikaev2013photonic}. In both of these scenarios, since TRS is preserved, the total Chern number is equal to zero, thus the eigenvalues are winding with opposite slopes. The shape of the Wilson loop is the same as in the case of a fragile phase, but being $\mathbb{Z}_2$-insulators strong topological phases, adding any new set of bands to the topological ones preserves the winding of the Wilson loop.

%%%%%%%%%%%%%%%%%%%%%%%%%%%%%%%%%%%%%%%%%%%%%%%%%%%%%%%%%%%%%%%%%%%%%%%%%%%%%%%%%%
%%%%%%%%%%%%%%%%%%%%%%%%%%%%%%%%%%%%%%%%%%%%%%%%%%%%%%%%%%%%%%%%%%%%%%%%%%%%%%%%%%

\section{Fundamentals of topology, the discrete limit}\label{TopologicalInvariantsDiscrete}
In this section we provide a practical method for evaluating the topological invariants introduced in the previous section. Usually, we do not have access to a continuous form of the periodic part of the field $u_{n,\mathbf{k}}(\mathbf{r})$, but only to its values on a set of momenta $\{\mathbf{k}_i\}$ in reciprocal space. We detail below a practical way to define topological invariants on a finite grid defined by the $\{\mathbf{k}_i\}$ momenta. Please note that in this section we are focusing on TE modes.

%%%%%%%%%%%%%%%%%%%%%%%%%%%%%%%%%%%%%%%%%%%%%%%%%%%%%%%%%%%%%%%%%%%%%%%%%%%%%%%%%%%%%
%
%
%%%%%%%%%%%%
\begin{figure}[ht]
\includegraphics[width=0.45\columnwidth]{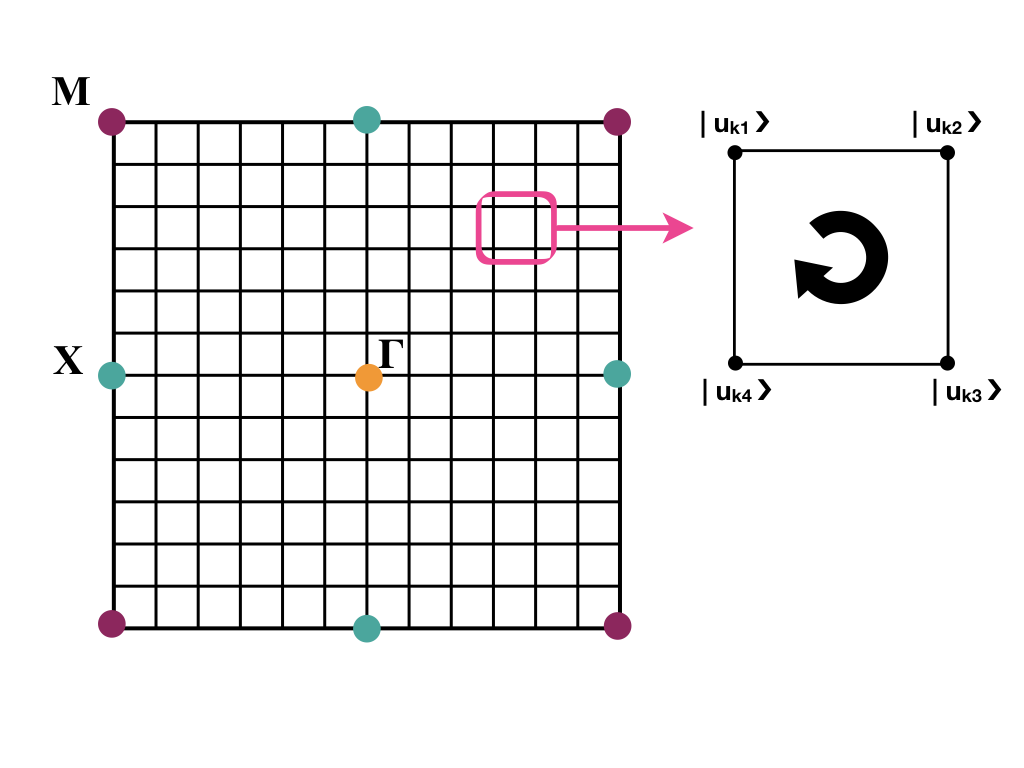}
\includegraphics[width=0.45\columnwidth]{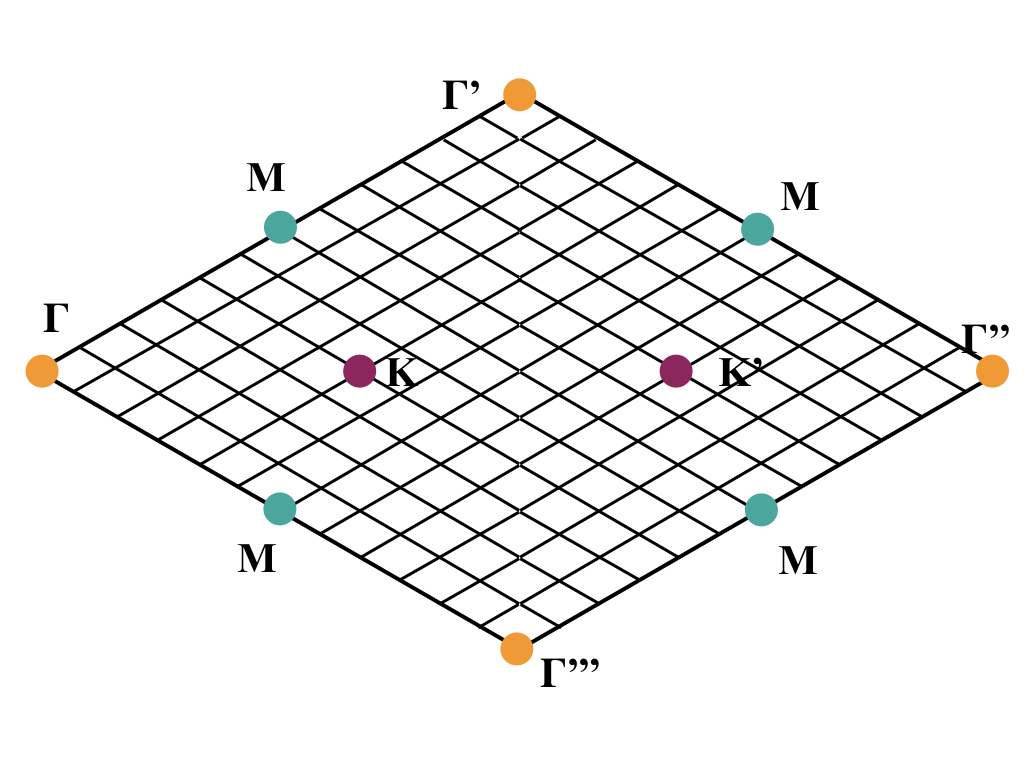}
\caption{ Discretization of the first BZ for the case of the square (left) and hexagonal (right) lattice. The position of the corresponding high symmetry points are indicated with colored points. In the inset we show a zoom-in of the plaquette over which the Berry curvature is evaluated --- plaquette of the discretization.}\label{fig:BZ}
\end{figure}
%%%%%%%%%
%
%
\subsection{Discretization of the first Brillouin zone}

As a first step, we need  to determine the unit cell and the corresponding point group symmetry of the 2D periodic system that we want to investigate. We can then define its first BZ and  discretize this cell into a regular mesh in $k$-space. Although in theory the number of points should not affect to the result of the calculation, we recommend a grid size at least of $24\times 24$ points for accurate description. We show in Fig.~\ref{fig:BZ} two examples of discretization of the first BZ of common systems: the square and hexagonal (or triangular) lattices. The choice of the BZ is not unique, in general it is better to choose the most convenient BZ to simplify the discretization: \emph{e.g.}, in the case of the triangular lattice (right panel of Fig.~\ref{fig:BZ}), instead of choosing a hexagonal BZ, it is easier to choose the rhombus defined by the reciprocal lattice vectors and discretize the cell in their directions.  
We recommend including enough points so that the high symmetry points in the Brillouin zone are included in the mesh. 
Finally, we compute numerically the Bloch eigenstates for each of the grid points, $\mathbf{k}_i$, in our case we solve the eigenvalue equation~\eqref{eq:master} using MPB~\cite{mpb}. 

%%%%%%%%%%%%%%%%%%%%%%%%%%%%%%%%%%%%%%%%%%%%%%%%%%%%%%%%%%%%%%%%%%%%%%%%%%%%%%%%%%%%%

\subsection{Discrete Berry curvature}\label{BERRY}                    

We use the four-point formula to compute the Berry curvature or Berry phase per unit cell for each plaquette of the discretization. For isolated bands, the Berry-Pancharatnam-Zak phase around a plaquette 
is given by~\cite{marzari1997maximally}:
%
%%%%%%%%%%%%%%%
\begin{align}\label{disBerry}
\phi= - \textrm{Im} ~ \log \Big[ \braket{u_{\mathbf{k}_{1}}(\mathbf{r})|u_{\mathbf{k}_{2}}(\mathbf{r})} \braket{u_{\mathbf{k}_{2}}(\mathbf{r})|u_{\mathbf{k}_{3}}(\mathbf{r})} \cdot \\ \braket{u_{\mathbf{k}_{3}}(\mathbf{r})|u_{\mathbf{k}_{4}}(\mathbf{r})} \braket{u_{\mathbf{k}_{4}}(\mathbf{r})|u_{\mathbf{k}_{1}}(\mathbf{r})}\Big] \nonumber\,,
\end{align}
%%%%%%%%%%%%%%%
%
%
where $u_{\mathbf{k}_{j}}(\mathbf{r})$ are the periodic functions at each corner of each plaquette (see Fig.~\ref{fig:BZ} --- central panel). Notice that along the closed path defined by each plaquette, the first and the last point coincide. 
Thus, any arbitrary phase coming from the diagonalization procedure cancels out since each state appears twice, once as a $\langle \mathrm{bra}|$ and once as a $|\mathrm{ket}\rangle$.

On the other hand, if there are one or more degeneracies in a group of $n$ bands, we need to replace the scalar products in Eq.~\eqref{disBerry} by the determinant of $(n\times n)$ overlap matrices. Equivalently, we can write:
%
%
%%%%%%%%%%%%%%%
\begin{equation}\label{disBerryDegeneracy}
    \phi = - \textrm{Im} ~ \log\Big\{  \det \left[\mathcal{S}_{\mathbf{k}_1 \mathbf{k}_2}\mathcal{S}_{\mathbf{k}_2 \mathbf{k}_3}\mathcal{S}_{\mathbf{k}_3 \mathbf{k}_4}\mathcal{S}_{\mathbf{k}_4 \mathbf{k}_1} \right] \Big\}
\end{equation}
%%%%%%%%%%%%%%%
%
%
\begin{widetext}
\noindent The overlap matrix, $\mathcal{S}$, between $\mathbf{k}$ and $\mathbf{k}'$ can be expressed as, %
%
%%%%%%%%%%%%%%%
\begin{equation}\label{overlap:matrix}
S_{\mathbf{k} \mathbf{k}'}=
\begin{bmatrix}
%  line 1
  \braket{u^{1}_{\mathbf{k}}(\mathbf{r})|u^{1}_{\mathbf{k}'}(\mathbf{r})} & \braket{u^{1}_{\mathbf{k}}(\mathbf{r})|u^{2}_{\mathbf{k}'}(\mathbf{r})} & 
  \ldots &
  \ldots &
  \braket{u^{1}_{\mathbf{k}}(\mathbf{r})|u^{n}_{\mathbf{k}'}(\mathbf{r})}\\
% line 2
   \braket{u^{2}_{\mathbf{k}}(\mathbf{r})|u^{1}_{\mathbf{k}'}(\mathbf{r})} & \braket{u^{2}_{\mathbf{k}}(\mathbf{r})|u^{2}_{\mathbf{k}'}(\mathbf{r})} & \braket{u^{2}_{\mathbf{k}}(\mathbf{r})|u^{3}_{\mathbf{k}'}(\mathbf{r})} &
   \ldots &
   \braket{u^{2}_{\mathbf{k}}(\mathbf{r})|u^{n}_{\mathbf{k}'}(\mathbf{r})}\\
% line 3
    \vdots &
    \braket{u^{3}_{\mathbf{k}}(\mathbf{r})|u^{2}_{\mathbf{k}'}(\mathbf{r})} & 
    \ddots & 
    \ddots &
    \vdots\\
% line 4
    \vdots &
    \ldots &
    \ddots &
    \ddots &
    \braket{u^{n-1}_{\mathbf{k}}(\mathbf{r})|u^{n}_{\mathbf{k}'}(\mathbf{r})} \\
% line 5
    \braket{u^{n}_{\mathbf{k}}(\mathbf{r})|u^{1}_{\mathbf{k}'}(\mathbf{r})} &
    \braket{u^{n}_{\mathbf{k}}(\mathbf{r})|u^{2}_{\mathbf{k}'}(\mathbf{r})} &
    \ldots &
    \braket{u^{n}_{\mathbf{k}}(\mathbf{r})|u^{n-1}_{\mathbf{k}'}(\mathbf{r})} &
    \braket{u^{n}_{\mathbf{k}}(\mathbf{r})|u^{n}_{\mathbf{k}'}(\mathbf{r})} 
\end{bmatrix}
\end{equation}
%%%%%%%%%%%%%%%
%
%
where the superscript $\ell$ of $u^\ell_{\mathbf{k}}$ indicates the band index. 
For TE modes, $u_\mathbf{k}(\mathbf{r})$ would be the periodic part of the eigenfunctions of the magnetic fields as defined in Eq.~\eqref{eq:1}.
Taking into account that $\mu=1$, the scalar products in the previous equations are defined as follows: 
%
%
%%%%%%%%%%%%%%%
\begin{equation}
    \braket{u_{\mathbf{k}}(\mathbf{r})|u_{\mathbf{k}'}(\mathbf{r})}= \sum_{i,j} u_{\mathbf{k}}^{*}(i,j) u_{\mathbf{k}'}(i,j) \Delta S 
\end{equation}
%%%%%%%%%%%%%%%
%
%
where $i$ and $j$ are the indices of the eigenvectors in the real-space, and $\Delta S$ is the differential of the surface.

If we consider TM modes instead, the solutions for the electric field would be simpler to manipulate, since they become scalar functions ($\mathbf{E}_\mathbf{k}(\mathbf{r})=E_\mathbf{k}(\mathbf{r})\hat{z}$). Defining the Bloch wavefunctions as
%
%
%%%%%%%%%%%%%%%
\begin{equation}
        E_{\mathbf{k}}(\mathbf{r})= e^{i\mathbf{k}\cdot\mathbf{r}} v_{\mathbf{k}}(\mathbf{r})\,,
\end{equation}
%%%%%%%%%%%%%%%
%
%
where $v_\mathbf{k}(r)$ is the periodic part of the electric field solutions, due to normalization constraints, the scalar products in Eqs.~(\ref{disBerry}-\ref{overlap:matrix}) would need to be replaced by products of the following form:
%
%
%%%%%%%%%%%%%%%
\begin{equation}
     \braket{v_{\mathbf{k}}(\mathbf{r})|v_{\mathbf{k}'}(\mathbf{r})}= \sum_{i,j}  [\mathbf{\varepsilon}(i,j)v_{\mathbf{k}}(i,j)]^{*}v_{\mathbf{k}'}(i,j) \Delta S\,.
\end{equation}
%%%%%%%%%%%%%%%
%
%
\end{widetext}
%%%%%%%%%%%%%%%%%%%%%%%%%%%%%%%%%%%%%%%%%%%%%%%%%%%%%%%%%%%%%%%%%%%%%%%%%%%%%%%%%%%%%

\subsection{Periodic gauge}
In this section we show how to fix the gauge choice for the periodic part of the field function. 
The eigenfunctions of a periodic system are described as Bloch states: 
%
%
%%%%%%%%%%%%%%%
 \begin{equation}\label{eq:1b}
    H_{\mathbf{k}}(\mathbf{r})= e^{i\mathbf{k}\cdot\mathbf{r}}\, u_{\mathbf{k}}(\mathbf{r})\,.
\end{equation}
%%%%%%%%%%%%%%%
%
%
In $\mathbf{k}$-space,  states translated by a reciprocal lattice vector $\mathbf{G}$ are equivalent:
%
%
%%%%%%%%%%%%%%%
\begin{equation}\label{eq:2}
    H_{\mathbf{k}+\mathbf{G}}(\mathbf{r})= e^{i(\mathbf{k}+\mathbf{G})\cdot\mathbf{r}}\, u_{\mathbf{k}+\mathbf{G}}(\mathbf{r})= H_{\mathbf{k}}(\mathbf{r})\,.
\end{equation}
%%%%%%%%%%%%%%%
%
%
Nevertheless, after diagonalization the eigenstates of the system gain a different arbitrary phase at each $\mathbf{k}$-point. 

Certain calculations, such as the Wilson loops, require that the condition of Eq.~\eqref{eq:2} is satisfied. Therefore, we need to fix the arbitrary phase of the boundary eigenstates using:
%
%
%%%%%%%%%%%%%%%
\begin{equation}\label{eq:4}
    u_{\mathbf{k}+\mathbf{G}}(\mathbf{r})= e^{-i\mathbf{G}\cdot\mathbf{r}} u_{\mathbf{k}}(\mathbf{r})\,,
\end{equation} 
%%%%%%%%%%%%%%%
%
%
we can express the phase factor of the previous expression as:
%
%
%%%%%%%%%%%%%%%
\begin{equation}\label{eq:5}
    e^{-i\mathbf{G}\cdot\mathbf{r}} = \frac{u_{\mathbf{k}+\mathbf{G}}(\mathbf{r})}{H_{\mathbf{k}+\mathbf{G}}(\mathbf{r})}  e^{i\mathbf{k}\cdot\mathbf{r}}\,,
\end{equation}
%%%%%%%%%%%%%%%
%
%
where we have used the formal boundary condition Eq.~\eqref{eq:2}. 
The phase factor on the right-hand side can be obtained from Eq.~\eqref{eq:1},
%
%
%%%%%%%%%%%%%%%
\begin{equation}\label{eq:6}
    e^{i\mathbf{k}\cdot\mathbf{r}} = \frac{H_{\mathbf{k}}(\mathbf{r})}{u_{\mathbf{k}}(\mathbf{r})}\,.
\end{equation}
%%%%%%%%%%%%%%%
%
%
Combining Eqs.~\eqref{eq:5} and \eqref{eq:6}, we obtain the phase factor in which we are interested 
%
%
%%%%%%%%%%%%%%%
\begin{equation}\label{eq:7}
    e^{-i\mathbf{G}\cdot\mathbf{r}} = \frac{u_{\mathbf{k}+\mathbf{G}}(\mathbf{r})}{H_{\mathbf{k}+\mathbf{G}}(\mathbf{r})}  \frac{H_{\mathbf{k}}(\mathbf{r})}{u_{\mathbf{k}}(\mathbf{r})}\,.
\end{equation}
%%%%%%%%%%%%%%%
%
%
Finally, we redefine the periodic part of the field at $(\mathbf{k}+\mathbf{G})$ by inserting the previous phase factor into Eq.~\eqref{eq:4}: 
%
%
%%%%%%%%%%%%%%%
\begin{equation}
    u'_{\mathbf{k}+\mathbf{G}}(\mathbf{r})= \frac{u_{\mathbf{k}+\mathbf{G}}(\mathbf{r})}{H_{\mathbf{k}+\mathbf{G}}(\mathbf{r})} H_{\mathbf{k}}(\mathbf{r})\,.
\end{equation}
%%%%%%%%%%%%%%%
%
%
where $u_{\mathbf{k}+\mathbf{G}}(\mathbf{r})$ is the periodic part of the field at $(\mathbf{k}+\mathbf{G})$ obtained directly from numerical evaluation --- in our case with MPB~\cite{mpb} --- and $H_{\mathbf{k}}(\mathbf{r})$ and $H_{\mathbf{k}+\mathbf{G}}(\mathbf{r})$ are the Bloch states at $\mathbf{k}$ and $(\mathbf{k}+\mathbf{G})$, respectively, also directly obtained from MPB~\cite{mpb}. \\
We have fixed the phase at the boundaries by implementing this new corrected periodic function $u'_{\mathbf{k}+\mathbf{G}}(\mathbf{r})$ along the boundaries of the first Brillouin zone, for all $(\mathbf{k}+\mathbf{G})$ positions.

Importantly, we have to be careful with the $\mathbf{\Gamma}$-point and apply the transformation individually otherwise the transformation will be applied twice. Moreover, we have to take into account that the energy of the first band of photonic system at $\mathbf{\Gamma}$-point usually is equal to zero. Thus, if we look at Eq.~\eqref{eq:TE}, at zero frequency any constant function is a solution to Maxwell's equation. Although MPB~\cite{mpb} gives a null eigenfunction as the default solution, we choose a constant, normalized eigenfunction to avoid singularities. 

%%%%%%%%%%%%%%%%%%%%%%%%%%%%%%%%%%%%%%%%%%%%%%%%%%%%%%%%%%%%%%%%%%%%%%%%%%%%%%%%%%%%%

\subsection{Chern and valley-Chern number}

The Chern number can be calculated over a discretized Brillouin zone by integrating the Berry curvature of each plaquette previously defined in Eqs.~\eqref{disBerry} and~\eqref{disBerryDegeneracy}:
%
%
%%%%%%%%%%%%%%%%
\begin{equation}\label{Chern:dis}
    C=  {{1}\over{2\pi}} \sum_{\text{BZ}} \phi_{j} = {{1}\over{2\pi}} \sum_{\text{BZ}} \text{Im} \left [ \log \prod _{i} \braket{u_{\mathbf{k}_{i}}|u_{\mathbf{k}_{i+1}}}  \right]\,,
\end{equation}
%%%%%%%%%%%%%%
%
%
The result of this calculation is an integer number. When TRS is preserved, this number is strictly zero. On the contrary, if TRS is broken, this number can take non-zero integer values. In those cases, we call the systems Chern insulators. If we consider a heterogeneous system composed of a finite size topological part ($C\neq 0$) and a trivial part ($C=0$), it will present edge states at the interface: the number of edge states is directly connected to the Chern number of the topological part --- the so-called bulk-edge correspondence~\cite{hasan2010colloquium}. 

On the other hand, if TRS is preserved and we consider pairs of high symmetry points related by TRS --- \emph{e.g.}, the $\mathbf{K}$ and $\mathbf{K}'$ points for a hexagonal or triangular lattice.
Although $C=0$, we can define a valley-Chern number dividing the first BZ in regions that enclose only one of the high symmetry points related by TRS operator, the computation procedure for the valley-Chern number is the same as the one employed in Eq.~\eqref{Chern:dis}, but reducing the integration of the Berry curvature to one half of the first BZ. In the case of a square lattice it  can be done with the same four-points formula, while for the hexagonal case, an additional set of plaquettes of three points must be added in order to divide the BZ in  two identical parts drawing the division line from $\mathbf{\Gamma}' $ to $\mathbf{\Gamma}''' $ (Fig.~\ref{fig:BZ} right panel). Therefore, this ``topological invariant'' will be half-integer with the opposite sign for each half of the first BZ.

%%%%%%%%%%%%%%%%%%%%%%%%%%%%%%%%%%%%%%%%%%%%%%%%%%%%%%%%%%%%%%%%%%%%%%%%%%%%%%%%%%%%%

\subsection{Discrete Wilson loop}

The calculation of the Wilson loops has to be performed differently for the case of an isolated band or for the case of a group of degenerate bands. For the former case, we can calculate a discrete Wilson loop in the following way
%
%
%%%%%%%%%%%%%%%
\begin{equation}\label{Wilson:one:discret}
    W(k_i)= - \text{Im}   \left( \log \prod_{k_j}\braket{u_{(k_i,k_j)}(\mathbf{r})|u_{(k_i,k_{j+1})}(\mathbf{r})} \right).
\end{equation}
%%%%%%%%%%%%%%%
%
%
This formula indicates that the Wilson loop can be computed for each $k_i$ in the $k_1$ direction by taking the phase of the final product of the overlap between pairs of consecutive periodic functions along $k_2$ direction. 
For the case of a group of degenerate bands, we replace the scalar product by a product of overlap matrices with the same structure as the ones described in Eq.~\eqref{overlap:matrix}, the Wilson loop then reads:
%
%
%%%%%%%%%%%%%%%
\begin{equation}
     W(k_i)= - \text{Im}\left[ \log \left( \prod_{k_j} S_{(k_i,k_j),(k_i,k_{j+1})}\ \right)\right].   
\end{equation}
%%%%%%%%%%%%%%%
%
%

In this case, we construct the overlap matrices between consecutive points along the lines in the $k_2$ direction. Then we multiply the overlap matrices for each pair of points element by element, and the resultant matrix must be diagonalized.
The phases of its eigenvalues then encode information about the position of the (hybrid) Wannier centers in real space along the $k_1$ direction. 

%%%%%%%%%%%%%%%%%%%%%%%%%%%%%%%%%%%%%%%%%%%%%%%%%%%%%%%%%%%%%%%%%%%%%%%%%%%%%%%%%%%%%
%%%%%%%%%%%%%%%%%%%%%%%%%%%%%%%%%%%%%%%%%%%%%%%%%%%%%%%%%%%%%%%%%%%%%%%%%%%%%%%%%%%%%

\section{Topological Photonic systems: examples}

In this section we apply the procedures described above to different topological systems and we explain each of the results. 

%%%%%%%%%%%%%%%%%%%%%%%%%%%%%%%%%%%%%%%%%%%%%%%%%%%%%%%%%%%%%%%%%%%%%%%%%%%%%%%%%%%%%

\subsection{Valley-Chern Insulator}

\begin{figure}[ht]
\includegraphics[width=1\columnwidth]{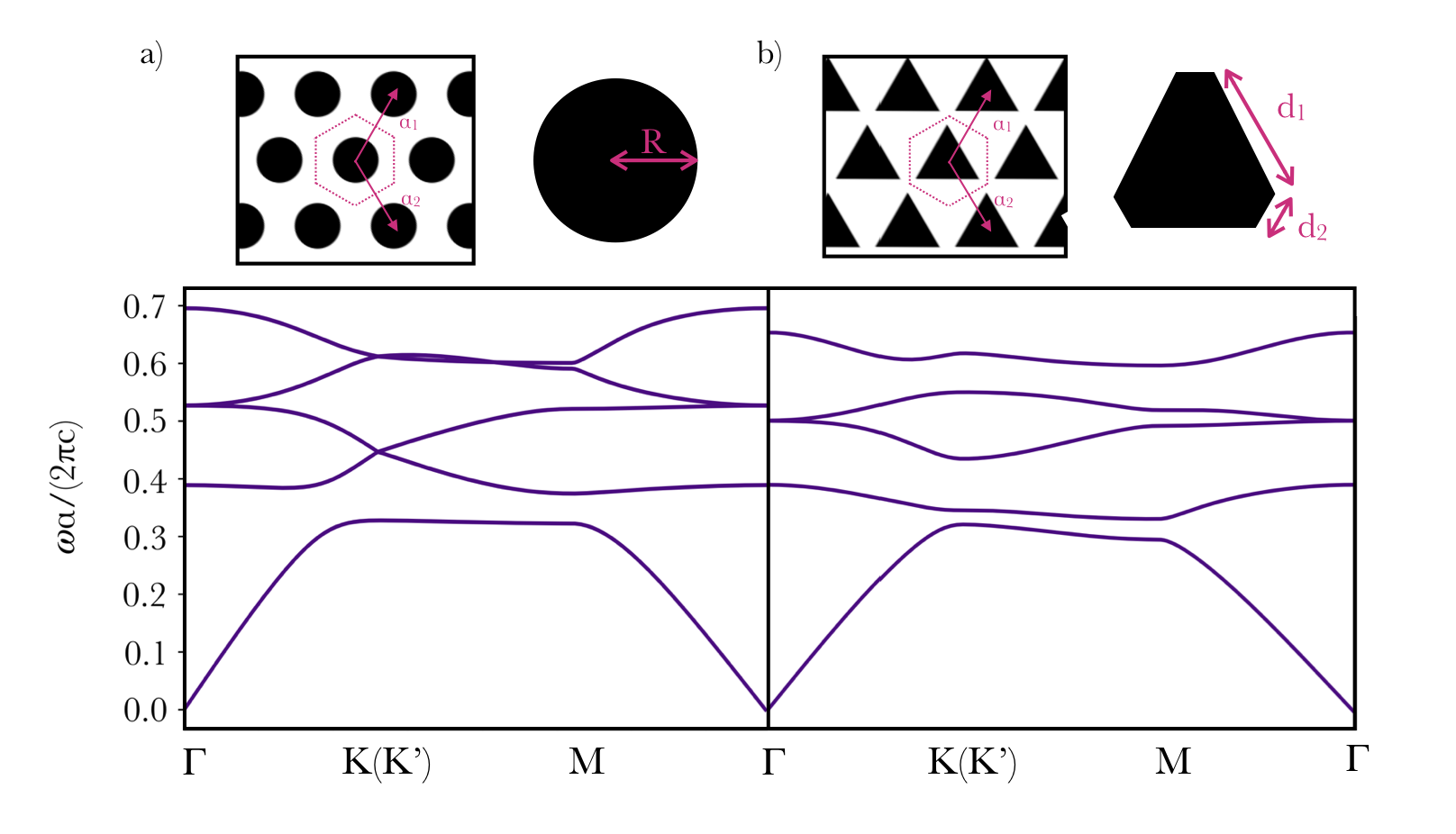}
\caption{a) TE band structure and geometry of triangular array of cylinders of radius radius $R=0.3075a_0$ and $\epsilon$=13. b) Band structure and geometry of triangular array of triangles with $\epsilon$=13. The length of the larger edge is $d_1= 0.825a_0$ and for the shorter one is $d_2= 0.055a_0$. Unitary lattice vectors for both are $a_1=(1/2, \sqrt{3}/2)a_0$ and $a_2=(1/2, -\sqrt{3}/2)a_0$.}\label{fig:valleybands} 
\end{figure}

We construct a system that possesses a valley degree of freedom. It is based on the one described by Ma and Shvets in Ref.~\cite{Ma2016}. Starting from a triangular array of dielectric cylinders made of silicon, $\epsilon$=13, and radius R=0.3075$a_0$ ---  upper panel of Fig.~\ref{fig:valleybands}a). The spectrum of the TE modes of the system presents a degeneracy point at $\mathbf{K}$ and $\mathbf{K}'$ between the second and the third band [lower panel of Fig.~~\ref{fig:valleybands}a)]. 
This degeneracy can be removed by reducing the symmetry of the system, for example by transforming the cylinders into cut triangles 
--- upper panel of Fig.~\ref{fig:valleybands}b). In the spectrum of this lattice, we can observe the lifting of the degeneracy at the $\mathbf{K}$-point between the second and the third band [lower panel of Fig.~~\ref{fig:valleybands}b)].

%
%
%%%%%%%%%%%%%%%
%%%%%%%%%%%%%%%
%
%

%
%
%%%%%%%%%%%%%%%
\begin{figure}[ht]
\includegraphics[width=1\columnwidth]{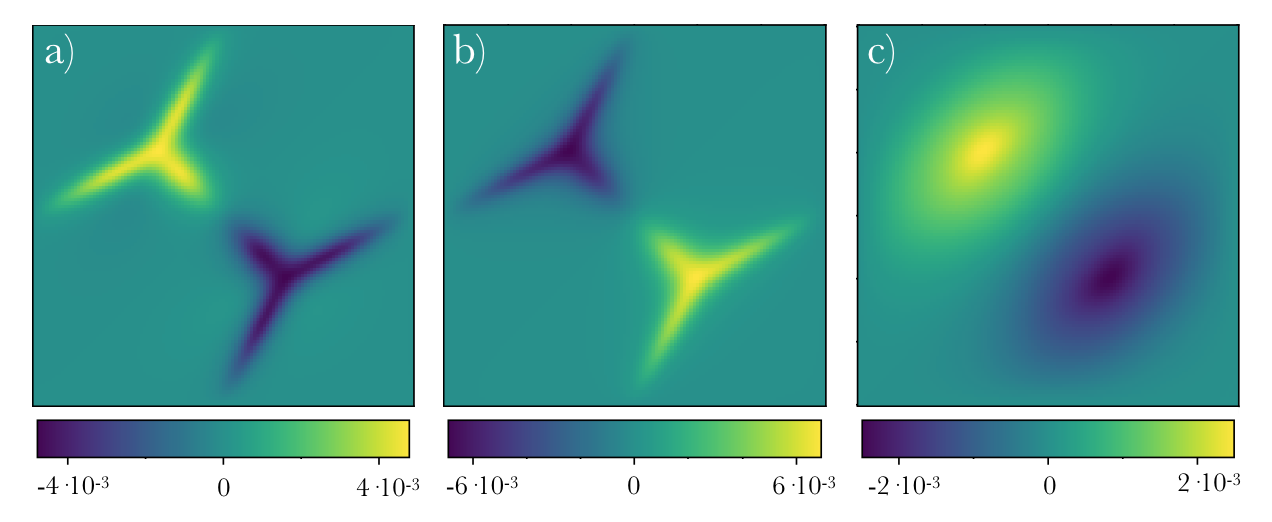}
\caption{Berry curvature distribution for the first BZ for the first three bands in the reciprocal space, computed with the four-point formula and plotted with squared shape. We can see that there is a discontinuity of this phase around the $\mathbf{K}$-point and the same but with opposite sign around its time-reversed partner, $\mathbf{K'}$-point.}\label{fig:triphase}
\end{figure}
%%%%%%%%%%%%%%%
%
%
The total Chern number associated to this gap is equal to zero, because TRS is preserved. By looking at the Berry curvature (see Fig.~\ref{fig:triphase}) of each band, we observe that its  spatial modulation is equal and opposite around the $\mathbf{K}$ and $\mathbf{K}'$ points . 
We can compute the valley-Chern number around the $\mathbf{K}$ and $\mathbf{K}'$ points, by integrating the Berry curvature over half of the BZ containing only one of the high symmetry points~\cite{Ezawa2013}. Around the $\mathbf{K}$ point the valley-Chern number for this  gap  --- associated with the second band --- is $C_v= +0.5$, and around $\mathbf{K}'$ is $C_v= -0.5$.

\begin{figure}[ht]
\includegraphics[width=1\columnwidth]{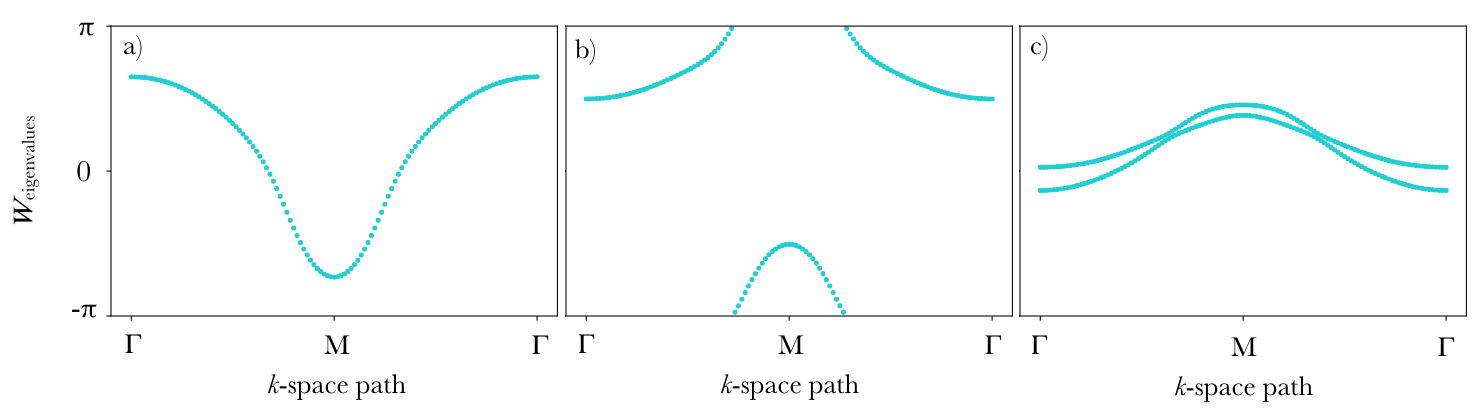}
\caption{Eigenvalues of the Wilson loop for the first band a), the second b) and the set of the following two degenerate bands, third and fourth c). All of them show trivial character due to the zero-valued Chern number.}\label{fig:WLvalley}
\end{figure}

The absence of a total Chern number is confirmed by the analysis of the Wilson loops, we can see the displacement of the Wannier centers but without any winding (Fig.~\ref{fig:WLvalley}). 
Since the rate of change of the Wilson loop is proportional to the Berry curvature, we can see also that these plots are consistent with the curvature maps (Fig.~\ref{fig:triphase}).
%
%
%%%%%%%%%%%%%%%
%%%%%%%%%%%%%%%
%
%

%%%%%%%%%%%%%%%%%%%%%%%%%%%%%%%%%%%%%%%%%%%%%%%%%%%%%%%%%%%%%%%%%%%%%%%%%%%%%%%%%%%%%

\subsection{\emph{Photonic} OAL Insulator}
In this section we explore the topological properties of one of the most popular designs employed in the field of topological photonics, introduced by Wu and Hu in Ref.~\cite{wuandhu,Wu2016}.

\begin{figure}[ht]
\includegraphics[width=1\columnwidth]{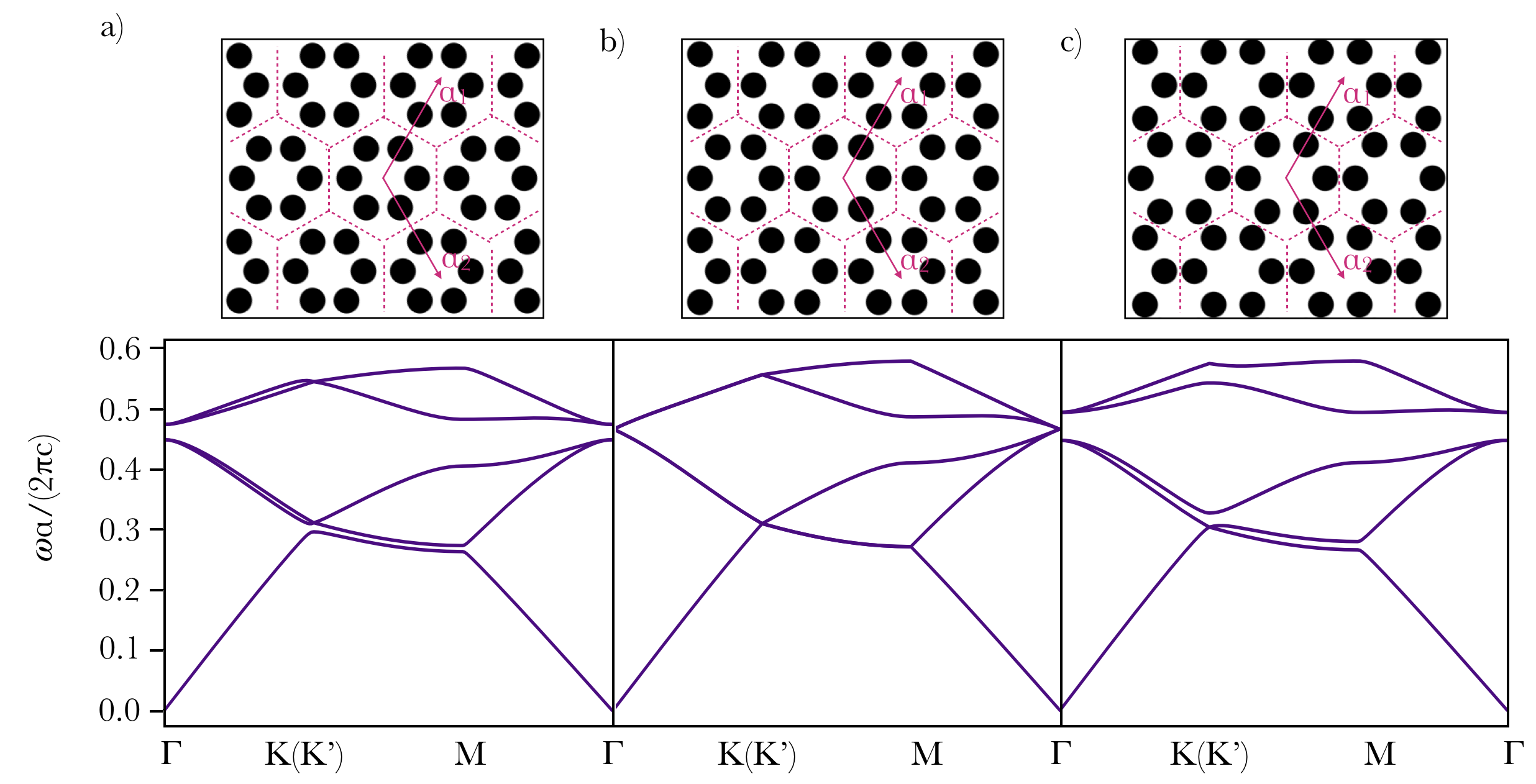}
 \caption{TM band structure of the artificial honeycomb lattice formed by six cylinders per unit cell. Panel b) refers to the cylinders placed at $1/3a_0$ from the center; a) refers to a contracted lattice, with the center of the cylinders moved to $1/3.16a_0$, and c) to an expanded lattice with the centers of the cylinders at $1/2.78a_0$. For the three of them the unitary vectors are $a_1=(1/2, \sqrt{3}/2)a_0$ and $a_2=(1/2, -\sqrt{3}/2)a_0$.}\label{fig:bandsOAL}
\end{figure}

The system is a triangular lattice with a unit cell containing six cylinders  ($\epsilon=11.7$) with radius $R= 0.12a_0$, placed at a distance $1/3a_0$ from the center of the unit cell. The band structure of this  lattice --- upper panel of Fig.~\ref{fig:bandsOAL} b) --- shows an artificial fourfold degeneracy at the $\mathbf{\Gamma}$-point emerging from the folding of the bands due to the choice of a nonprimitive (larger) unit cell. This enlarged unit cell is necessary because later the rods are displaced from their original position. 
%
%
%%%%%%%%%%%%%%%
%%%%%%%%%%%%%%%
%
%

\begin{figure}[ht]
\includegraphics[width=1\columnwidth]{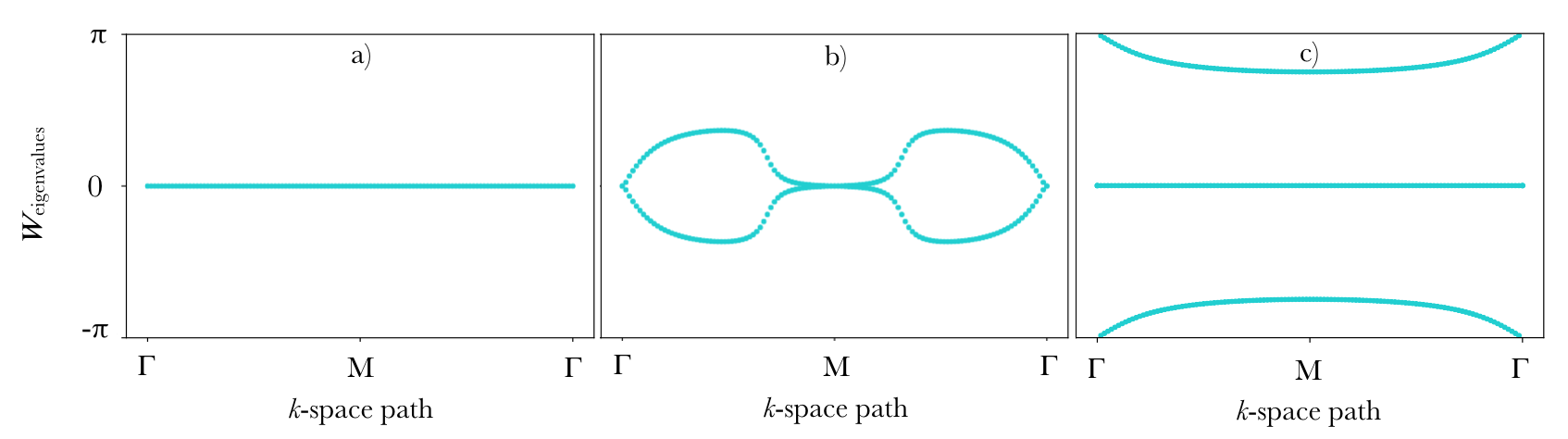}
\caption{Wilson loops of the contracted a) and b) and expanded c) artificial honeycomb lattice, respectively. The contracted lattice shows trivial Wilson loops while the expanded shows the characteristic Wilson loop of an OAL.}\label{fig:WLOAL}
\end{figure}

Displacing the cylinders to different distances from the center of the unit cell these artificial degeneracies are lifted and topological properties can emerge in the associated gaps. Placing the cylinders at $a_0/3.16$, contracting the lattice --- upper panel of Fig.~\ref{fig:bandsOAL} a) --- the artificial degeneracies are broken leaving the first band isolated, and two sets of degenerate bands: one group of bands formed by the second and the third band and, another set formed by the fourth and the fifth. Both sets of bands present degeneracies at $\mathbf{\Gamma}$ and $\mathbf{K}$. To analyze the topological character of those new gaps we can look at the Wilson loop --- Fig.~\ref{fig:WLOAL} a) and~\ref{fig:WLOAL} b). The Wilson loop of the first band has a constant value equal to zero, which is representative of a trivial gap. For the set of the second and third bands we can see that the projected position of the Wannier centers are slightly moving but without any winding, preserving the trivial character. 
On the other hand, if the cylinders are placed at $a_0/2.78$, expanding the lattice --- upper panel of Fig.~\ref{fig:bandsOAL} c)--- only a gap between the third and the fourth band is opened. In this case there is a crossing at $\mathbf{K}$ between the first and the second band and another cross at $\mathbf{\Gamma}$ between the second and the third, therefore to study the topological character of this new gap we have to study the three lowest bands together. Looking at the Wilson loop --- Fig.~\ref{fig:WLOAL}c) --- we observe no windings but it becomes apparent that the Wannier centers are also localized at the edge of the unit cell ($W=\pm\pi$), instead of only in the center ($W=0$) as in the trivial case, indicating that the system presents an obstruction similar to that emergent in the Su-Schrieffer-Heeger  chain in 1D~\cite{vanderbilt2018}. 
In analogy to condensed matter, we call optical systems presenting this phenomenology \emph{photonic} OAL insulators. 
%
%
%%%%%%%%%%%%%%%
%%%%%%%%%%%%%%%
%
%

%%%%%%%%%%%%%%%%%%%%%%%%%%%%%%%%%%%%%%%%%%%%%%%%%%%%%%%%%%%%%%%%%%%%%%%%%%%%%%%%%%%%%

\subsection{Fragile-Phase Insulator}
\begin{figure}[ht]
\includegraphics[width=1\columnwidth]{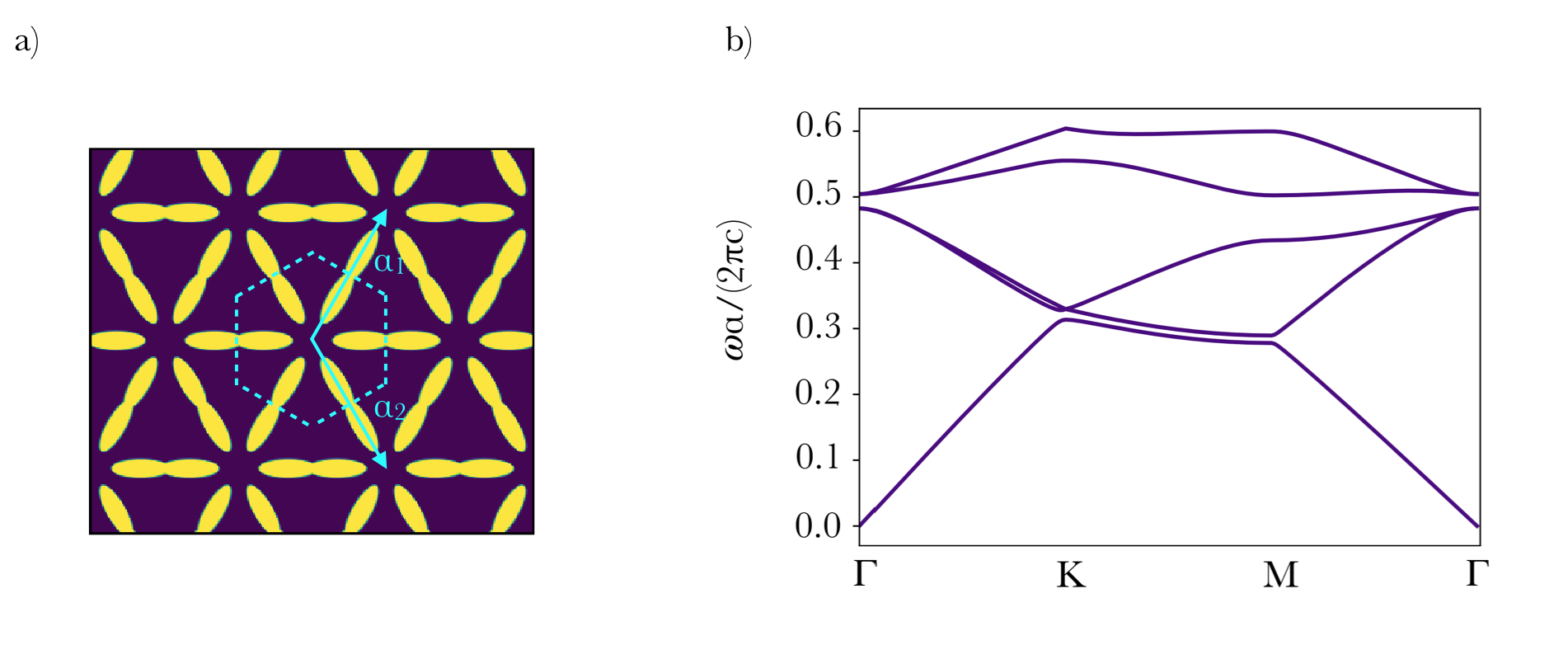}
\caption{a) Schematic representation of the system consisting on a triangular array with $a_1=(1/2, \sqrt{3}/2)a_0$ and $a_2=(1/2, -\sqrt{3}/2)a_0$, composed by ellipses whose diagonals length are $d_1=0.4a_0$ and $d_2=0.13a_0$ with $\epsilon=11.7$ and placed at $a_0/3$. b) The corresponding TM band dispersion.}\label{fig:Bandsfragile} 
\end{figure}

\begin{figure}[ht]
\includegraphics[width=1\columnwidth]{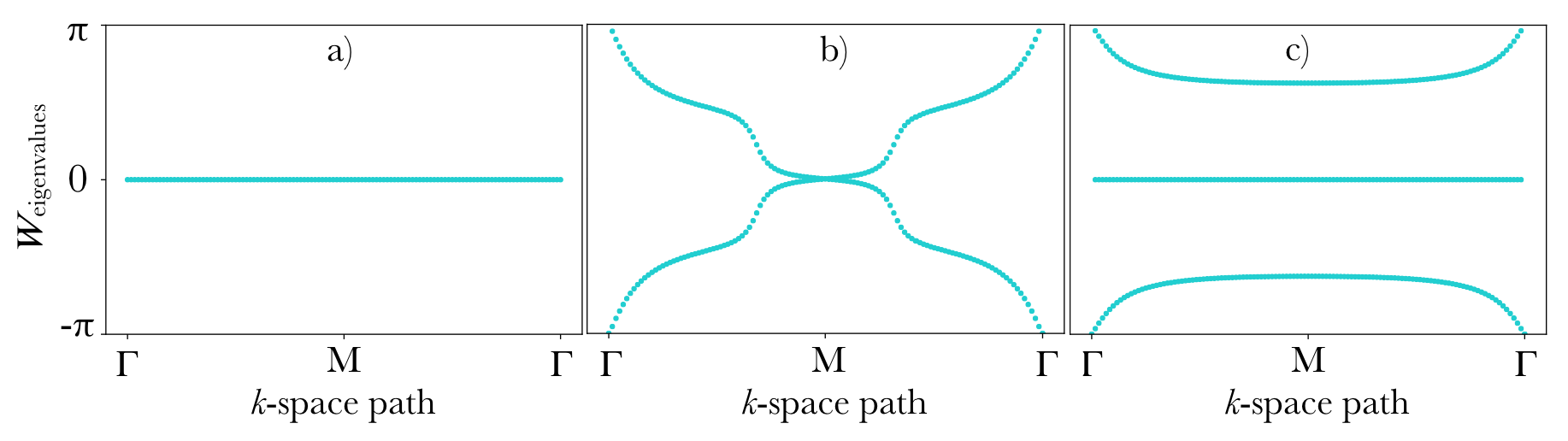}
\caption{Wilson loop for: a) first band that possesses trivial topological character, b) set of the second and third band which has two windings with opposite sign and c) set of the three lowest bands that present the shape of an OAL, proving the topological fragility of the former set.}\label{fig:fragile}
\end{figure}

To show a fragile photonic phase, we use the system described in \cite{autocita}. Each unit cell of the system consists of an array of six dielectric ellipses of silicon ($\epsilon=11.7$) whose length of the two principal axes are $d_1=0.4a_0$ and $d_2=0.13a_0$, and placed at $a_0/3$ from the center of the unit cell --- Fig.~\ref{fig:Bandsfragile}a). The corresponding band dispersion --- Fig.~\ref{fig:Bandsfragile}b) --- shows two main gaps: one between the first and the second band and another between the third and the fourth. The first band is isolated while the second and the third present degeneracies at $\mathbf{K}$ and $\mathbf{\Gamma}$. The variation of the length of the ellipses' axes drives a topological phase transition, from fragile to trivial, and from trivial to OAL, by closing and reopening gaps. For more details see Ref.~\cite{autocita}. 
%
%
%%%%%%%%%%%%%%%
%%%%%%%%%%%%%%%
%
%
We analyze the Wilson loops to identify the topology of this phase. For the first band --- Fig.~\ref{fig:fragile}a) --- the Wilson loop has a constant value equal to zero which means that the Wannier centers are localized in the center of the unit cell. This shape is typically associated with trivial isolated bands. For the set of the second and the third bands, we see that the Wilson loops wind with opposite signs. Therefore, the total Chern number for this set of bands is equal to zero. This shape of the Wilson loop is reminiscent of $\mathbb{Z}_2$ insulators, and occurs frequently for fragile bands with twofold rotational symmetry~\cite{bradlyn2019disconnected}. 
The difference between both is that the $\mathbb{Z}_2$ preserves the winding if a trivial set of bands is added to the calculation, while the winding determining the fragile invariant is destroyed by the addition of a trivial band. Thus, we have to analyze a new subset of bands, the one that contains the three lowest bands. In this case, the winding of the Wilson loop is destroyed and takes the shape of an OAL (Fig.~\ref{fig:fragile}c). This is a clear proof of the topological fragility associated to the Wilson loop of the second and third bands. 
%
%
%%%%%%%%%%%%%%%
%%%%%%%%%%%%%%%
%
%

%%%%%%%%%%%%%%%%%%%%%%%%%%%%%%%%%%%%%%%%%%%%%%%%%%%%%%%%%%%%%%%%%%%%%%%%%%%%%%%%%%%%%

\subsection{Chern Insulator}\label{Chernexample}

We conclude this tutorial presenting the case of a photonic crystal where both the electric permittivity $\epsilon\to\epsilon(\mathbf{r})$ and the magnetic permeability $\mu\to\mu(\mathbf{r})$ are periodic functions and can in general be non-diagonal tensors. This particular example corresponds to the case of the gyromagnetic Chern insulator proposed by Wang \emph{et al.}~\cite{Wang2008}. \\

\begin{figure}[ht]
\includegraphics[width=1\columnwidth]{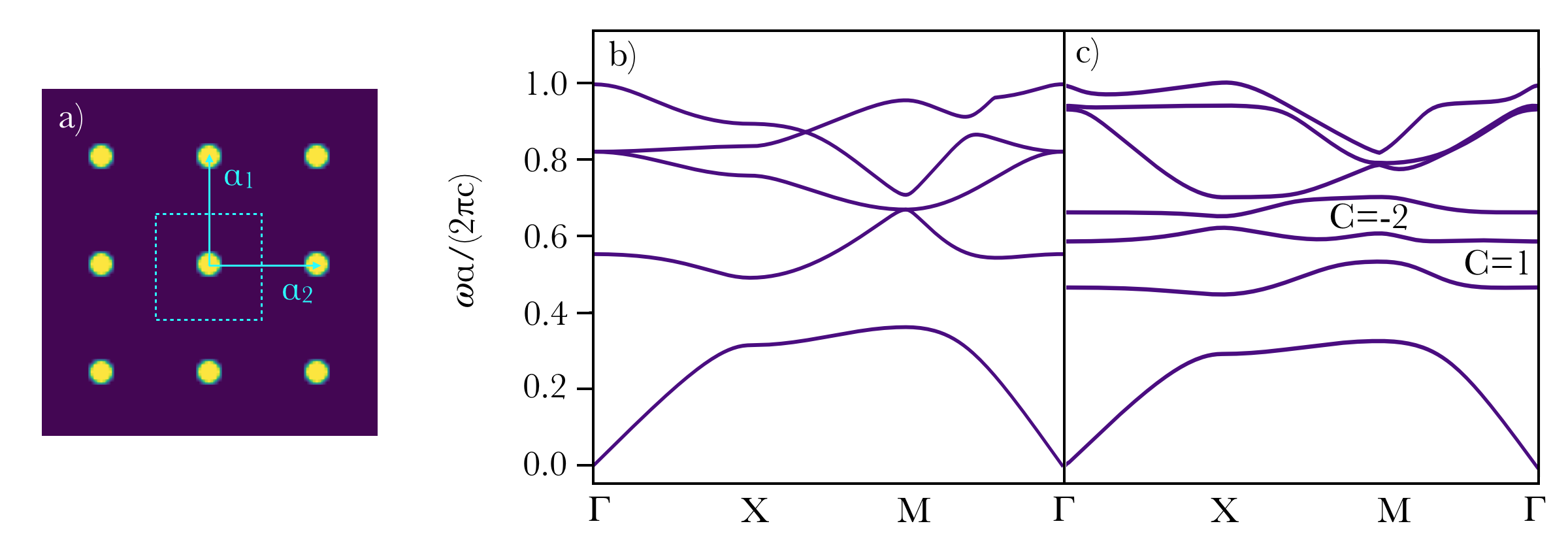}
\caption{a) Square array of YIG magneto-optic cylinders of radius $R=0.11a_0$ in air. The two translational vectors are: $a_1=(1,0)a_0$ and $a_2=(0,1)a_0$. TM band structure for the trivial system without magnetic field b) and for the Chern insulator with applied external magnetic field c).}\label{fig:BANDSCHERN} 
\end{figure}

The system consists of a 2D photonic crystal composed of a square array of yttrium-iron-garnet (YIG) magneto-optic cylinders in air with radius R=0.11$a_0$ and  $\epsilon=15$ [see Fig.~\ref{fig:BANDSCHERN}a)]. When an external magnetic field is applied in $z$-direction, it induces a gyromagnetic anisotropy resulting in the following permeability tensor inside each of the rods:
%
%
%%%%%%%%%%%%%%%
\begin{equation}\label{eq:mu:tensor}
\bm{\mu}=\begin{bmatrix}
  \mu & i\kappa & 0 \\
   -i\kappa & \mu & 0 \\
    0 & 0 & \mu_0 \\
\end{bmatrix},
\end{equation}
%%%%%%%%%%%%%%%
%
%
with  $\mu=14\mu_0$ and $\kappa=12.4\mu_0$. According to Ref.~\cite{Wang2008}, these values correspond to an applied magnetic field of $0.16$~T. The corresponding equation that describes the propagation of TM modes in the system reads:
%
%
%%%%%%%%%
\begin{equation}\label{eq:mu:epsilon}
\nabla\times \left[\frac{1}{\mu(\mathbf{r})} \nabla\times E_z(\mathbf{r}) \right] = \left(\frac{\omega}{c}\right)^2\epsilon(\mathbf{r})E_z(\mathbf{r})\,.
\end{equation}
%%%%%%%%%
%
%
This equation is a generalized eigenvalue problem. 

Assuming that $E_z(\mathbf{r})=e^{i\mathbf{k}\cdot\mathbf{r}} v_\mathbf{k}(\mathbf{r})$, the correct normalization of the eigensolutions requires the use the following scalar product in the calculation of the topological invariants:
%
%
%%%%%%%%%%%%%%%
\begin{equation}
    \braket{v_{\mathbf{k}}(\mathbf{r})|v_{\mathbf{k}'}(\mathbf{r})}= \sum_{i,j} [\epsilon(i,j) v_{\mathbf{k}}(i,j)]^{*} v_{\mathbf{k}'}(i,j) \Delta S
\end{equation}
%%%%%%%%%%%%%%%
%
Moreover, if the use of magnetic field solutions is preferred, then defining $\mathbf{H}(\mathbf{r})=e^{i\mathbf{k}\cdot\mathbf{r}} \mathbf{u}_\mathbf{k}(\mathbf{r})$, and the correct normalization of the eigensolutions requires the use the following scalar products
%
%
%%%%%%%%%%%%%
\begin{equation}
        \braket{\mathbf{u}_{\mathbf{k}}(\mathbf{r})|\mathbf{u}_{\mathbf{k}'}(\mathbf{r})} = \sum_{i,j} \mathbf{u}_{\mathbf{k}}(i,j)\cdot\bm{\mu}(i,j)\mathbf{u}_{\mathbf{k}'}(i,j) \Delta S,
\end{equation}
%%%%%%%%%%%
%
%
where $\cdot$ denotes the scalar product in the two-dimensional polarization space:
%
%
%%%%%%%%%%%%%%%
\begin{eqnarray}
    \mathbf{u}\cdot \mathbf{u'}=u^{*}_{x}u'_{x}+u^{*}_{y}u'_{y}.
\end{eqnarray}
%%%%%%%%%%%%%%
%
%

These ideas have been experimentally realized  in the same set-up but employing different gyromagnetic  materials~\cite{wang2009observation}. The experiments have shown the presence of uni-directional edge states associated to the quantized Chern numbers of the system between the second and the third band. A similar experiment in the same set-up has been also conducted as a function of the applied magnetic field to show further topological features of the system~\cite{Skirlo2015}.

In Panels b) and c) of Fig.~\ref{fig:BANDSCHERN}, we show the band structure for the trivial and the topological case, respectively. The transition between the two cases is obtained by switching on the external magnetic field. For the trivial case, we  observe degeneracy  between the second and the third band at $\mathbf{M}$-point and between the third and the fourth at $\mathbf{\Gamma}$-point. The only gap in the system is between the first and the second band. This gap is preserved when the magentic field is switched on, however the Chern number associated with the first band is zero. On the other hand, the magnetic field removes the degeneracies of the higher bands, resulting in the emergence of topological properties. The Chern number associated to the bands from the second to the fourth are different from zero and have the values shown in the Panel c) of Fig.~\ref{fig:BANDSCHERN}. 

%
%
%%%%%%%%%%%%%%%
\begin{figure}[ht]
\includegraphics[width=1\columnwidth]{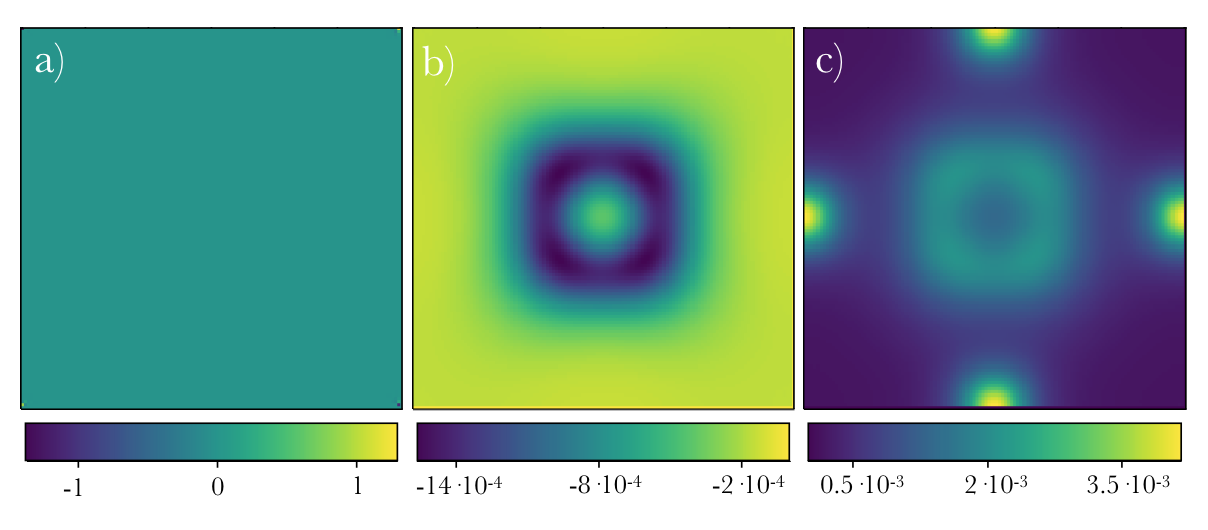}
\caption{Berry phase distribution in the unit cell for each band. a) The trivial character of the first band is evidenced by a constant Berry phase equal to zero, while the higher bands -- the second b) and third c) -- present a phase structure since the Chern number is different from zero.}\label{fig:phases} 
\end{figure}

\begin{figure}[ht]
\includegraphics[width=1\columnwidth]{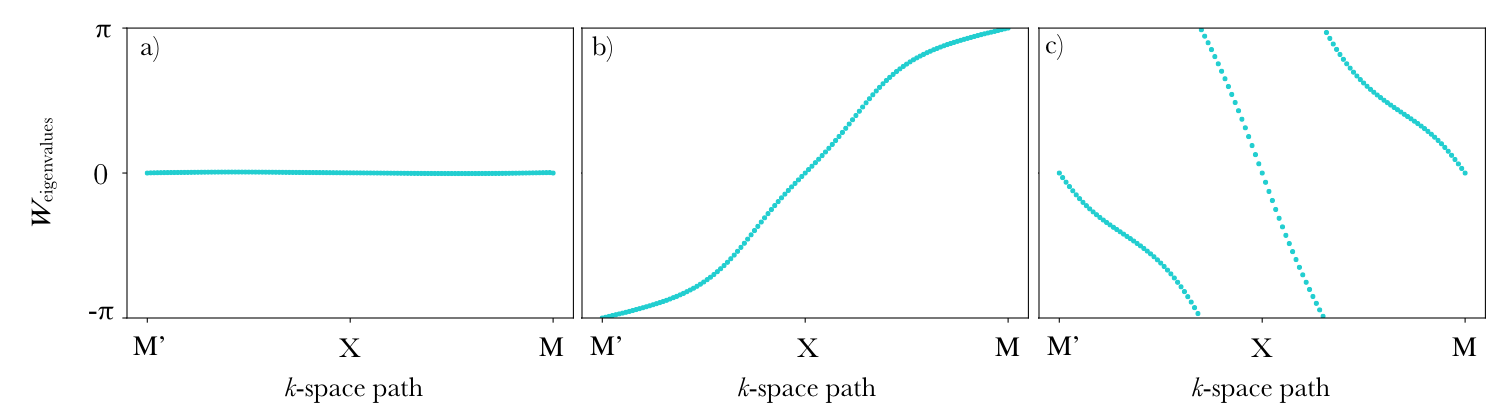}
\caption{Wilson loop for each band of the system. (a) Trivial band, $C=0$, (b) second band, $C=1$ and (c) third band, $C=-2$.}\label{fig:WL}
\end{figure}

%%%%%%%%%%%%%%%
%
%
We extract the topological information of the gyromagnetic 
curvature for the first four bands using the methods described before.
As we can see in Fig.~\ref{fig:phases}a), the the Berry phase of the first band is homogeneous and equal to zero, thus corresponding to a trivial band, while for the rest of the bands Figs.~\ref{fig:phases}b)-d), the Berry phase presents a spatial modulation  with a finite average value. This modulation of the phase indicates that as the eigenvector is propagating in a closed path, there is a manifestation of topological character for those bands due to the application of the external magnetic field. It means that those bands possess an associated Chern number different from zero.
%
%
%%%%%%%%%%%%%%%
%%%%%%%%%%%%%%%
%
%
These results are further confirmed by the analysis of the Wilson loops. We can observe that the first band is trivial, Fig.~\ref{fig:WL}a). The Wilson loop of the second band winds with positive slope, meaning that the Chern number of this band is $1$. For the third band we observe two windings of the Wilson loop eigenvalue, in this case with positive slope. Therefore the Chern number of this band is $-2$. Please note that the slope of the Wilson loop depends on the sign convention chosen in the exponential in Eq.~\eqref{3.7}.
%
%
%%%%%%%%%%%%%%%
%%%%%%%%%%%%%%%
%
%

%%%%%%%%%%%%%%%%%%%%%%%%%%%%%%%%%%%%%%%%%%%%%%%%%%%%%%%%%%%%%%%%%%%%%%%%%%%%%%%%%%%%%
%%%%%%%%%%%%%%%%%%%%%%%%%%%%%%%%%%%%%%%%%%%%%%%%%%%%%%%%%%%%%%%%%%%%%%%%%%%%%%%%%%%%%

\section{Conclusion \& Outlook}

This paper is a tutorial for the characterization of the topological properties of 2D photonic crystals. 
In particular, it explains how to compute the Berry phase and curvature for photonic crystal structures over a discretized reciprocal space. By integrating these quantities over the appropriate regions of the BZ, these methods facilitate the numerical calculation of topological invariants such as the Chern and valley-Chern number in photonic systems.  
Moreover, we provide a numerical method to calculate Wilson loops in photonic crystals, a very useful tool for determining the topological properties both in systems preserving TRS and with broken TRS. 
Additionally, to help the reader to be familiar with the practical realization of topological photonic system, we have shown examples of different phases: a system with valley degree of freedom, OAL, fragile phase and a Chern insulator. Explaining the main characteristics of each phase and how to interpret the results obtain to classify each topological phase.

The methods described here could be generalized in several interesting
directions: including losses and pumping leads to the field on
non-Hermitian materials, which comprises interesting new physics, \emph{e.g.}, breaking of time-reversal invariance and the appearance of exceptional
points and non-equilibrium steady states. This would entail generalizing
the program to include non-orthogonal Bloch states~\cite{Shen_Zhen_Fu_2018,Zhao2019}.

Photonic crystals made on \emph{non-linear} materials --- and including
squeezing terms at the quantum level --- can give rise to new topological
phases distinct from those obtained in condensed-matter~\cite{Peano_Houde_Brendel_Marquardt_Clerk_2016,Shi_Kimble_Cirac_2017}.

We have focused on solutions to Maxwell's equation that do not propagate
in the $z$-direction --- translationally invariant. Allowing for
$k_z\not=0$, topological properties of propagating solutions in arrays
of optical waveguides~\cite{Szameit_Nolte_2010}, can be studied.

%%%%%%%%%%%%%%%%%%%%%%%%%%%%%%%%%%%%%%%%%%%%%%%%%%%%%%%%%%%%%%%%%%%%%%%%%%%%%%%%%%%%%
%%%%%%%%%%%%%%%%%%%%%%%%%%%%%%%%%%%%%%%%%%%%%%%%%%%%%%%%%%%%%%%%%%%%%%%%%%%%%%%%%%%%%
\begin{acknowledgments}
\section*{Acknowledgement}
MBP thanks M. Olano for fruitful  discussions. MGV and AGE acknowledge the IS2016-75862-P and FIS2016-80174-P national projects of the Spanish MINECO, respectively and DFG INCIEN2019-000356 from Gipuzkoako Foru Aldundia. AGE received funding from the Fellows Gipuzkoa fellowship of the Gipuzkoako Foru Aldundia through FEDER "Una Manera de hacer Europa", and by Eusko Jaurlaritza, grant numbers, KK-2017/00089, IT1164-19, and KK-2019/00101. The work of MBP, GG, JJS, DB and AGE is supported by the Basque Government through the SOPhoQua project (Grant PI2016-41). The work of DB is supported by Spanish Ministerio de Ciencia, Innovation y Universidades (MICINN) under the project FIS2017-82804-P, and by the Transnational Common Laboratory \emph{QuantumChemPhys}.
\end{acknowledgments}

\end{document}